\documentclass[preprint]{aastex}

\newcommand{\Msun}{\ensuremath{{\rm M}_{\sun}}}
\newcommand{\Lsun}{\ensuremath{{\rm L_{\sun}}}}
\newcommand{\Rsun}{\ensuremath{{\rm R_{\sun}}}}

\newcommand{\myemail}{bdeupree@ap.smu.ca}

\shorttitle{The Structure of Close Binaries in 2D}

\shortauthors{Deupree \& Karakas}

\begin{document}


\title{The Structure of Close Binaries in Two Dimensions}


\author{R. G. Deupree}
\affil{Institute for Computational Astrophysics, Department of Astronomy \& Physics,
Saint Mary's University, Halifax, NS, B3H 3C3, Canada}
\email{\myemail}

\and

\author{A. I. Karakas}
\affil{Institute for Computational Astrophysics, Department of Astronomy \& Physics,
Saint Mary's University, Halifax, NS, B3H 3C3, Canada}
\email{akarakas@ap.smu.ca}




\begin{abstract}

The structure and evolution of close binary stars has been studied 
using the two-dimensional (2D) stellar structure algorithm developed
by \citet{deupree95}. We have calculated a series of solar composition stellar 
evolution sequences of binary models, where the mass of the 2D model is 8$\Msun$
with a point-mass 5$\Msun$ companion.  We have also studied the structure
of the companion in 2D, by considering the zero-age main-sequence (ZAMS) structure 
of a 5$\Msun$ model with an 8$\Msun$ point-mass companion. This result suggests 
that treating the 5$\Msun$ star as a point source for the 8$\Msun$ evolution 
is reasonable. In all cases the binary orbit was assumed to be circular and 
co-rotating with the rotation rate of the stars. We considered binary 
models with three different initial separations, $a = 10, 14$ and 20$\Rsun$. 

These models were evolved through central hydrogen burning or until the 
more massive star expanded to fill its critical potential surface or
Roche lobe. The model with a separation of 20$\Rsun$ will be expected to
go through Case B type mass transfer, during the shell H-burning phase.
The 14$\Rsun$ model is expected to go through mass transfer much
earlier, near the middle of core hydrogen burning, and the 10$\Rsun$ model
is very close to this situation at the ZAMS. 

The calculations show that evolution of the deep interior quantities is only 
slightly modified from those of single star evolution. Describing the model 
surface as a Roche equipotential is also satisfactory until very close to the 
time of Roche lobe overflow, when the self gravity of the model about to lose
mass develops a noticeable aspherical component and the surface time scale
becomes sufficiently short that it is conceivable that the actual surface 
is not an equipotential.

\end{abstract}




\section{Introduction}




Much progress has been made understanding the structure and evolution of
single stars. This is because we assume that the stars are spherically
symmetric and in or close to hydrostatic equilibrium for most of their evolution, 
allowing the use of one dimensional models. Observations have shown that there are
places where these assumptions do not hold, for example in rapidly rotating
main-sequence stars such as the Be star Achernar \citep*{desouza03,jackson04,lovekin05}.
Binary (or multiple) star evolution on the 
other hand, is far more complex because of the possibility of the two components
interacting, which depends on the mass of the stars and the orbital parameters 
of the system.  However, it is essential that the evolution of multiple stellar
systems be studied given that most stars in the solar neighbourhood are observed
to be in such systems. Many of the interacting stars are observed to be 
undergoing mass transfer via Roche Lobe Overflow (RLOF) which results in a 
veritable zoo of stellar subtypes including the W Ursa Majoris contact 
binaries \citep*[][and references therein]{li04} and the Algol-type binaries 
\citep{nelson01} which are formed by mass transfer on or near the main sequence.
Other subtypes, such as cataclysmic variables, novae and symbiotics \citep{it96}, 
X-ray binaries \citep{verbunt93} and black hole X-ray novae \citep[for example][]{mc01} 
are all formed through one or more mass transfer event when the stars are in the 
final stages of stellar evolution.  The recent high quality X-ray observations 
have been providing important links between some of these subtypes, including 
classical and recurrent novae experiencing X-ray emission, symbiotics and cataclysmic 
variables \citep{prp02,wheatley03}.   Millisecond pulsars are also an important class of objects 
that are formed when mass is transfered from a companion onto an old neutron star,
resulting in the spin-up of the neutron star \citep{pk94}. 
Stellar nucleosynthesis and galactic chemical evolution also depend on the 
interactions between binary stars to produce
explosive events such as classical novae or Type Ia supernovae, where asymmetrical 
effects play an important role \citep{t04,yl04}.
Recent work suggests that the presence of a binary companions may also be important
for Type II supernova explosions \citep{pod04}. 

The pioneering studies by \cite{kw67} and \citet{pacz66,pacz67a,pacz67b} and
\citet{pacz71}
set the stage for much of the work that followed on close binary evolution,
which by definition involves systems in which the two stars are close enough to
allow at least one phase of mass transfer via RLOF. These early studies 
focused on upper main sequence stars that were evolved through either
classical Case A mass transfer, which takes place during the main sequence or
Case B mass transfer which takes place after the exhaustion of 
central hydrogen \citep{kw67}. Whilst the input physics and the stellar 
models have undergone significant improvements in recent years, the 
assumptions made when calculating binary evolution remain essentially 
the same as those used in these early studies 
\citep[for example,][]{nelson01}: that a star in a binary 
system is treated as spherically symmetric, even if it fills its Roche lobe, that 
the mass transfer from a star that overfills its Roche lobe is treated as 
spherically symmetric, that the matter which leaves the mass-losing star is 
accreted in a spherically symmetric manner on the surface of the mass-gaining 
star, that the orbit is circular (though see \citet*{regos05} for a relaxation
of this assumption), that the star is assumed to be in hydrostatic
equilibrium, that there is a critical radius $R_{\rm L}$ such that mass exchange 
takes place when $R \ge R_{\rm L}$ and that the radius of a star is always 
equal to or less than $R_{\rm L}$. The total mass of the system and the angular
orbital momentum are also conserved during the evolution. 
In this study, we are concerned with the first assumption, i.e. how valid is the
assumption of spherical symmetry up to the beginning of mass transfer. 
The key points of the one dimensional approach are that the interior is 
unaffected by the presence of the companion, while the surface is defined
as a Roche potential surface for two gravitational point sources.

Whilst the proper way to treat this problem is in three dimensions, we can
adequately describe the problem in two dimensions (2D) if we assume that the orbit 
is circular and co-rotating with the spins of the two stars. We have
used the 2D structure algorithm developed by \citet{deupree90,deupree95} to
study the structure and evolution of intermediate mass solar composition stars
with a point mass companion. The models we consider are of a similar mass and 
composition studied by Paczynski in his early series of papers on the evolution 
of close binary evolution \citep{pacz66,pacz67a,pacz67b}, which assumed a mass 
of 8$\Msun$ for the more massive component, hereafter the primary,
and 5.3$\Msun$ for the least massive component, hereafter the secondary.
The aim of this paper is to study the evolution and structure
of an 8$\Msun$ primary with a 5$\Msun$ point mass companion, from
the zero aged main sequence (ZAMS) through to the end of central H-burning
or to the point where mass transfer is expected to begin. We investigate
varying the separation between the stars, from 10$\Rsun$ to 20$\Rsun$
(corresponding to periods of 1.01 to 2.86 days respectively). The results
from these stellar models are examined to see if they satisfy the
constraints used in studies of close binary evolution, namely does the
primary star remain in or close to spherical symmetry prior to RLOF?
We have also studied the ZAMS structure of a 5$\Msun$ model with an
8$\Msun$ point source to see if the assumption of treating the secondary
star as a point mass source is reasonable.

The paper is organized as follows.
The numerical method is presented in \S\ref{section:numerical} along with
details of the stellar models calculated. In  \S\ref{section:results} we
present our results, and we finish with a discussion and
some conclusions in \S\ref{section:discussion}.

\section{Models} \label{section:numerical}

The stellar models were calculated with the 2D stellar structure, evolution, and 
hydrodynamics code developed by \citet{deupree90,deupree95}. The algorithm uses a 2D finite 
difference approach to solve the appropriate equations 
via a fully implicit Henyey \citep*{HFG} technique. More strictly speaking, 
the calculations are 2.5D because the azimuthal velocity is calculated even though azimuthal 
symmetry is imposed.  The independent variables are the fractional surface radius, the 
spherical polar coordinate ($\theta$), and time.
The equations to be solved simultaneously in the implicit formulation are the 
conservation laws for mass, three components of momentum, energy,
and hydrogen abundance, along with Poisson's equation.
The dependent variables are the temperature, density, three components of velocity, 
hydrogen mass fraction, and the gravitational potential. The unknown surface radius is 
determined from the equation that the integral of the density distribution over the volume 
equals the total mass. The usual subsidiary relations are satisfied with the composite 
hydrogen burning nuclear energy generation rates \citep{FCZ67}, the OPAL equation of 
state \citep{RSI96}, and OPAL radiative opacity \citep{IG96}. 
The latter two are included as tables, with values and numerical derivatives brought 
into the Jacobian needed for the Henyey iteration scheme as required.

Because the independent variables are not Lagrangian variables, we must calculate the velocities
to determine how the material moves with respect to the coordinate system. Thus we must retain
all advective and time dependent terms in the equations. 
Unlike our approach to rotating stars \citep{deupree90,lovekin05},
in which the calculations are performed in the inertial frame, here the calculations are
performed in a coordinate system which is rotating at a constant rate. The rotation rate is
given by the orbital period of the two bodies tidally locked in a circular orbit. This rate is
not allowed to change in the calculations presented here, which covers the time from the ZAMS
up to just before Roche lobe overflow, which is justified since no mass is allowed to be lost 
from the system (via a stellar wind for example) and hence the angular momentum is constant.   

We must prescribe the shape of the surface. For rotating stars this is done by assuming the
centrifugal acceleration can be described by a potential, even when this may not be true. In the
binary star calculations presented here we will force the surface to be an equipotential
whose three components are the self gravity of the model being calculated,
the point source gravitational potential of the companion, and the potential arising from the
centrifugal acceleration of the rotating coordinate system.
The surface boundary condition for the self gravitational
potential is calculated for each angle on the surface of a sphere just exterior to the largest
radius of the model. This condition is given by a weighted integral of the density
distribution over the volume of the model, and these integrals are included as part of the 
Henyey iteration scheme.

The convective core is taken to be adiabatic. No convective overshooting is included in these
calculations. There are no convective envelopes in the models computed here, so no recourse
to a non-adiabatic convection theory is required. The surface temperature is taken to be the
effective temperature divided by the fourth root of two, and the radiation of the companion on
the surface of the primary is neglected, the implications of which are discussed further in
\S\ref{section:discussion}.

The geometry of the calculations is that the model is symmetric about the axis defined by the
centers of the two stars. In the star we are modelling, we take the value of the spherical polar
coordinate, $\theta$ to be 0 degrees in the direction opposite the companion and 180 degrees in the 
direction toward the companion.  Thus, our model is a prolate spheroid.
Also, the rotation axis is at $\theta = 90\,$degrees at the azimuthal angle so that the 
rotation axis is perpendicular to the orbital plane.  The rotation
imposes symmetry about the rotation axis (which is perpendicular to the line between the
stellar centers), so that the model really should be an ellipsoid 
instead of a spheroid. However, even at close separations of the two stars the centrifugal
acceleration is sufficiently small in comparison with the gravitational acceleration of the
companion that a spheroidal shape is not a bad approximation, and that is adopted here.

We shall be addressing three specific configurations. The primary will be an 8$\Msun$ model and the
secondary will have 5$\Msun$. We will only be considering evolution prior to mass transfer
so these designations are unambiguous. The three cases we consider are separations between the stellar
centers of 10, 14, and 20$\Rsun$. The primary in the first case will fill its Roche lobe very soon
after the ZAMS and in the third case will fill its Roche lobe relatively early in hydrogen shell burning.
The filling of the Roche lobe by the primary in the second case occurs when nearly half of the 
hydrogen in the core has been burned. 
We will also compare the evolution of these primaries with that of a spherical
single star performed with the same radial and angular zoning. In these models we have 450
radial zones for the longest radius and twenty angular zones.
Most of our calculations will focus on the evolution
of the primary, but we will also examine the ZAMS model of the secondary (in the presence of a
point source potential with the primary's mass) to obtain some estimate as to how realistic
a point source potential for the secondary in the primary evolution calculations might be.
 
\section{Results} \label{section:results}

The assessment of the applicability of the 1D approach has two components: the interior
evolution and the surface characteristics.    To this end we have evolved
three 8$\Msun$ models from the ZAMS until just before Roche overflow begins. We stop the
calculations at this point because the 2D surface boundary conditions would no longer be
adequate once the component being modeled begins to lose mass to the other binary member.

We first examine the central conditions of the model with the greatest separation in 
Fig.~\ref{fig1}, a plot of the central temperature versus the central density.
The differences in the central temperature for a given central density between the binary
and the single star models for this separation are about 0.3 per cent, 
a number which is reasonably constant throughout
the entire main sequence evolution. The difference in percentage terms is approximately 
double this amount for the intermediate separation case and about 1.8 per cent for the 
smallest separation.   Because this last case
is already very close to Roche lobe overflow on the ZAMS, it is clear that the 
percentage cannot be much higher than this.

The mass of the convective core is plotted as a function of the central hydrogen
abundance in Fig.~\ref{fig2} for the intermediate separation case. Here we see that this
is unaffected by the presence of the binary companion as well. However,
it is true that the shape of the convective core boundary is not spherical,
but rather is elongated slightly in the direction of the binary companion. 
The elongation is only a fraction of a radial zone in the largest separation case,
a bit more than a zone in the intermediate separation, and about two zones
in the small separation case. At the convective core boundary location the zone size
is about 0.004 of the surface radius and a radial zone contains about 
0.1 $\Msun$ when integrated over all angles.
From these considerations it appears that the interior quantities
are not much affected by the presence of the companion.

We now turn to the surface configuration. We shall frame our discussion in
terms of the elongation of the model. The elongation is defined as the ratio
of the surface radius in the direction opposite the binary companion to the
surface radius in the direction of the binary companion. When the ratio is
above about 0.9, we find that the Roche equipotential fits the surface shape
as accurately as we can measure with our discrete zoning (approximately 0.4 per cent). 
As the elongation approaches 0.8 the two shapes are slightly different, mostly
in the direction between the two components of the binary system.
This is shown in Fig.~\ref{fig3}, a plot of the surface shape for the largest
separation model just before the gravitational contraction phase begins.
The difference in surface shape between the ROTORC and Roche models becomes
more significant as Roche lobe overflow is approached. A comparison for a model
very close to Roche lobe overflow (elongation of about 0.69) for the
intermediate separation case is shown in Fig.~\ref{fig4}. There is clearly a difference
in the two potentials in the direction between the two binary components. To be
fair the Roche potential contours are changing quite rapidly here in the
direction of looking more like the ROTORC contours for slightly larger fractional radii.

Clearly one factor in the differing potentials is the possibility that the
self gravity of the primary at the model surface is not that given by the
point source potential. The self gravitational potential of the primary on a
spherical surface whose radius is given by the largest surface radius in the
model is shown in Fig.~\ref{fig5}. Clearly the magnitude of the potential is largest
in the direction between the two components, and the amplitude variation is a
little more than one percent. Another way of viewing this is to examine
what we refer to as the ``column mass". This is defined by integrating the
radial density distribution at a given angle over a spherical volume element;
i.e., it is the interior mass distribution the model would have if this radial
distribution was spherically symmetric. We show this column mass for three
angles versus radial zone number in Fig.~\ref{fig6}, where it is evident that there
is little difference in the mass distribution interior to approximately 5$\Msun$.
Closer to the model surface there is more mass concentrated in the direction
toward the companion, and the column mass monotonically decreases going away
from this direction.

One interesting feature related to the shape of the surface is the rate of change in
that shape as the time of Roche lobe overflow is approached. We compare the
expansion rate of the surface in the direction of the companion for the largest
separation model with the surface expansion of a spherical model in Fig.~\ref{fig7}. 
Clearly, the primary surface in the direction of the secondary is expanding at a 
much larger rate, one that is fairly close to the expansion experienced in the 
early phases of hydrogen-shell burning. The expansion rate increases yet faster as 
Roche lobe overflow is approached, as is evidenced in Fig.~\ref{fig8}. Here we illustrate 
the expansion rate of the surface of the primary in the direction of the secondary 
in the intermediate separation case. The expansion rate is sufficiently rapid that 
it is possible that the stellar surface does not remain an equipotential during this 
phase although our calculations assume that is does.
This rapid expansion of the surface may also affect the rate of transfer 
from the primary to the rest of the system, at least near the start of mass transfer.

Another interesting feature is the distribution of the luminosity on the stellar
surface. The surfaces of constant temperature are all extended in the direction
of the secondary so that the direction of radiative diffusion is away from the line
between the centers of the two components. Thus, the radial flux drops
significantly as one approaches the surface from the interior in that direction. 
The radial flux drops in the opposite direction for the same reason, but to a 
much smaller extent.  This can be seen in Fig.~\ref{fig9}, a plot of the local 
``luminosity", defined to be the area of a spherical surface multiplied by the 
local radial flux, as a function of $\theta$. We present this luminosity at a 
radial distance about halfway from the model
center to the surface in the direction toward the companion and at the surface.
The trend mentioned is present at the deeper location, but is quite pronounced
at the surface. The small scale variations in the surface luminosity are produced by the
surface changing its radial zone number on this particular time step. The
total luminosity emitted through both surfaces is nearly the same, with the small
difference produced by the expansion of the outer layers of the model.

The enlarged surface area produced by the elongation of the model towards the
companion has the effect of reducing the average effective temperature, as seen
in the evolutionary tracks for a spherical model and the model with the largest
separation in Fig.~\ref{fig10}. This is somewhat illusionary, however, for it ignores
both the shine of the companion of the secondary onto the surface of the primary
and the fact that the observed effective temperature now becomes a function
of the location of the observer with respect to the system geometry.

In all this work we have assumed that the secondary may be treated by a
point source potential. It is of some interest to see how significant
this assumption is. Therefore we calculated the ZAMS model of the
secondary assuming the primary was the point source companion. This is
not completely self consistent, but it should be adequate to determine
some level of credibility for the assumption. We find that the self
gravity of the 5$\Msun$ ZAMS model with the 8$\Msun$ companion on
a spherical surface just exterior to the largest surface radius
varies by about 0.6 per cent on that surface. 

\section{Discussion and Conclusions} \label{section:discussion}

We have performed 2D stellar evolution sequences of the primary member of a binary
system. These sequences were carried through core hydrogen burning up to
the time of Roche lobe overflow. The interior stellar evolution characteristics
were altered from the single star evolution only on the fraction of a percent level 
even for the smallest separations. The mass of the convective core as a 
function of time is also unchanged,
although the convective core boundary is no longer a completely spherical surface. 
Thus, the approximation that the interior structure and evolution is the same as that 
for a single star appears to be quite good throughout this phase. This can probably 
be interpreted as being adequate for any stars which are not compact objects because 
the smallest separation case
was about as small as we could make it on the ZAMS for these two masses. The differences
in structure become noticeable at about one quarter of the radius (equivalent to about
an interior mass of 5$\Msun$) for the model just prior to Roche lobe overflow.

The surface is reasonably well approximated by the Roche surface until very close to the
beginning of Roche lobe overflow. As Roche lobe overflow is approached the contour lines
in the direction of the secondary become very sensitively dependent on the value of the
equipotential, and the variations between the Roche potential and the equipotential
determined from the models reflect this. It is also true that the timescale for the
surface change becomes very short as Roche lobe overflow is approached, raising the
possibility that the assumption that the surface is an equipotential is not valid. It
is difficult to identify any consequence of this from the standpoint of the evolution
calculations, but it may play a role in the details of the mass transfer within the
system.

One important assumption we make in the calculations is that we neglect the radiation
from the secondary component, and this radiation could make a considerable difference
to the effective temperature and observed surface flux of the primary star, 
particularly in the direction of the secondary.   We can estimate the
``true'' effective temperature of the primary star by calculating the effective temperature
at the surface of the primary using the secondary's luminosity and 
$L_{\rm S} = 4 \pi \sigma T_{\rm eff}^{4} R_{\rm D}$, where $L_{\rm S}$ is the ZAMS
luminosity of the 5$\Msun$ model determined by the 2D evolution code, $\sigma$ the 
Stefan-Boltzmann constant, $R_{\rm D}$ the distance from the center of the secondary to the
surface of the primary and $T_{\rm eff}$ the effective temperature. If we take the
$a = 14\Rsun$ case as a representative example and the radius of the primary just before
RLOF, which is 7.114$\Rsun$, then $R_{\rm D} = 6.886\Rsun$ (where the ZAMS radius of the
5$\Msun$ 2D model is 2.785$\Rsun$). Using $L_{\rm S} = 529\Lsun$ we find that the
effective temperature at the surface of the primary using the secondary's luminosity to
be 10561K. At the angular zone closest to the secondary, the effective temperature of the
primary just before RLOF was determined to be 13476K, which is 27\% hotter than 
the effective temperature calculated using the luminosity of the secondary model.

We have also calculated the effective temperature at the surface of the primary
using the secondary's luminosity just after the ZAMS. This effective temperature was 
found to be about 2.3 times less than the effective temperature of the primary 
calculated by the 2D evolution code. 
The implication of this result for the models near RLOF is that the surface 
flux of the primary, for example as shown in Fig. 9, should clearly be higher at the 
angular zones closest to the companion. This also means that the observed dip 
in the flux is not completely real and indeed, the effective temperature of the primary 
closest to the companion should be substantially hotter than what we have calculated 
it to be using the 2D stellar evolution code.  What effect the secondary's radiation has on 
the interior properties of the 8$\Msun$ primary are yet to be determined and we
leave that for future work. We do note that this effect will mostly be important when
the primary is rapidly approaching RLOF and the evolution of the system at this stage 
is quick enough that the radiation from the secondary should not alter our overall
conclusions.




\acknowledgments

Support for this work is provided by NSERC, the Canada Foundation for 
Innovation (CFI), the Nova Scotia Research Innovation Trust (NSRIT), and 
the Canada Research Chair (CRC) program 
for which the authors are grateful.



\appendix





\clearpage


\centerline{\bf FIGURE CAPTIONS}

\figcaption[fig1]{\label{fig1} Plot of the central temperature versus 
the central density for the evolution of a single 8$\Msun$ model (solid) and an 
8$\Msun$ model with a 5$\Msun$ companion with a separation between the centers of 
the models of 20$\Rsun$ (dash). The difference in the central densities for a given
central temperature is about 0.3 per cent and is nearly constant throughout
the solution.}

\figcaption[fig2]{\label{fig2} Plot of the mass of the convective core versus the central
hydrogen mass fraction for the evolution of a single 8$\Msun$ model (solid)
and an 8$\Msun$ model with a 5$\Msun$ companion with a separation between
the centers of the models of 14$\Rsun$ (dash). The mass of the convective
core is essentially unaffected by the presence of a companion even though
the shape of the convective core in the binary case is not perfectly spherical.}

\figcaption[fig3]{\label{fig3} Comparison of the surface shape of the 8$\Msun$ 
primary with a 5$\Msun$ secondary as determined by the 2D stellar evolution code 
(circles) and by the Roche potential surface (pluses). The location of the secondary
is below the bottom of the graph centered at $y$-coordinate of $-$2.54. The
solid curve is a circle to help highlight the departures from a spherical
surface. The two surfaces are quite close except near the line between the
centers.}

\figcaption[fig4]{\label{fig4} Comparison of the surface shape of the 8$\Msun$ 
primary with a 5$\Msun$ secondary as determined by the 2D stellar evolution code 
(circles) and by the Roche potential surface (dashed curve) for the intermediate
separation case. The location of the secondary is below the bottom of the
graph centered at $y$-coordinate $-$1.97. This model is very close to Roche
lobe overflow. The solid curve is a circle to highlight the departures from
a spherical surface. Note that the Roche potential surface and the 2D
stellar evolution code surfaces differ close to the line between the two
models. Part of this is just the sensitivity to the shape of the Roche
potential surface to the exact value of the contour as Roche lobe overflow
is approached.}

\figcaption[fig5]{\label{fig5} Plot of the self gravitational potential of 
the 8$\Msun$ primary on a spherical surface just exterior to the largest 
radius of the model versus the spherical polar coordinate theta. This is 
for the model very close to Roche lobe overflow presented in Fig.~\ref{fig4}.
Note  that the self gravity on this spherical surface varies by about one percent 
over the surface and is largest in amplitude in the direction towards the 
secondary.}

\figcaption[fig6]{\label{fig6} The interior column mass versus radial zone 
number in the direction away from the secondary (lower solid curve), 
perpendicular to the line joining the centers of the primary and secondary 
(upper solid curve), and in the direction towards the secondary (dashed curve). 
The interior column mass is calculated by taking the radial density 
distribution at a given angle and calculating the interior mass as if this 
density distribution were spherically symmetric. This presentation is for the
intermediate separation model just before Roche lobe overflow shown in the
previous two figures. The implication is that the mass of the primary has
been slightly redistributed by the presence of the secondary to be larger
in the direction of the secondary, but only outside the core of the
primary. This picture is consistent with the properties of the interior
parts of the star which control the evolution being only very slightly
affected by the presence of the secondary.}

\figcaption[fig7]{\label{fig7} Plot of the expansion velocity of the surface 
radius as a function of the central hydrogen mass fraction. The velocity for 
a spherical model is given by the solid curve, while the expansion velocity 
for the surface radius in the direction of the secondary for the primary 
binary component is given by the dashed curve. The model is the binary with 
the largest separation. The expansion of the primary accelerates in this 
direction as Roche lobe overflow is approached. This model gets close to Roche lobe
overflow before the end of core hydrogen burning, but then moves away from
it during the global gravitational contraction phase, and eventually
overflows the Roche lobe early in hydrogen shell burning.}

\figcaption[fig8]{\label{fig8} Plot of the expansion velocity of the surface 
radius as a function of central hydrogen mass fraction. The solid curve is 
for a single star model and the dashed curve for the intermediate separation 
case. Roche lobe overflow is expected to occur shortly after the last model 
shown in the dashed curve. Because the time scale during this end phase becomes 
very short, it may be possible that the surface is not an equipotential in this
latter phase.}

\figcaption[fig9]{\label{fig9} Plot of the product of the area of a spherical 
surface and the radial component of the radiative flux as a function of spherical 
polar coordinate theta. The solid curve is at a radius a little more than half
the way from the center to the surface in the direction of the binary
companion. The dashed curve is at the model surface. The radiation flow in
the regions between those given by the two curves is away from the line
between the centers of the two binary components, as one would expect from
the shape of the equipotential surfaces (see Fig.~\ref{fig3}). Note that any
radiation of the secondary onto the primary has been neglected.}

\figcaption[fig10]{\label{fig10} HR diagram of the evolution of a single star 
(solid) and of the primary member of a the binary system with the largest 
separation (dash).  The two models have the same luminosity history, as one 
would expect from the nearly identical deep interior structures, but the 
average effective temperature of the binary member is less as one might 
anticipate from Fig.~\ref{fig9}.  The actual effective temperature one would 
observe would depend on the orientation of the observer to the binary system.}

\clearpage 



\begin{figure}
\plotone{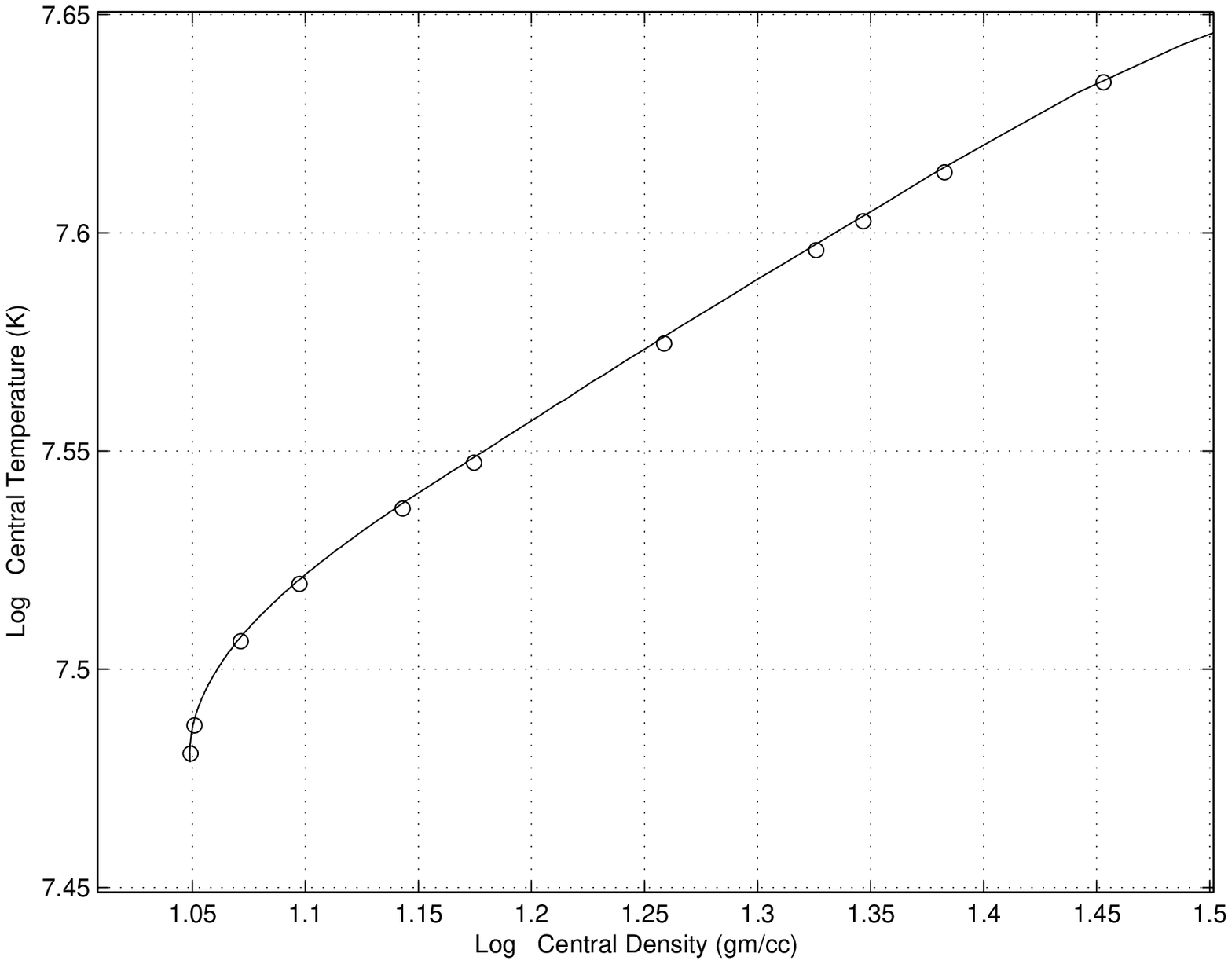}
\end{figure}

\clearpage

\begin{figure}
\plotone{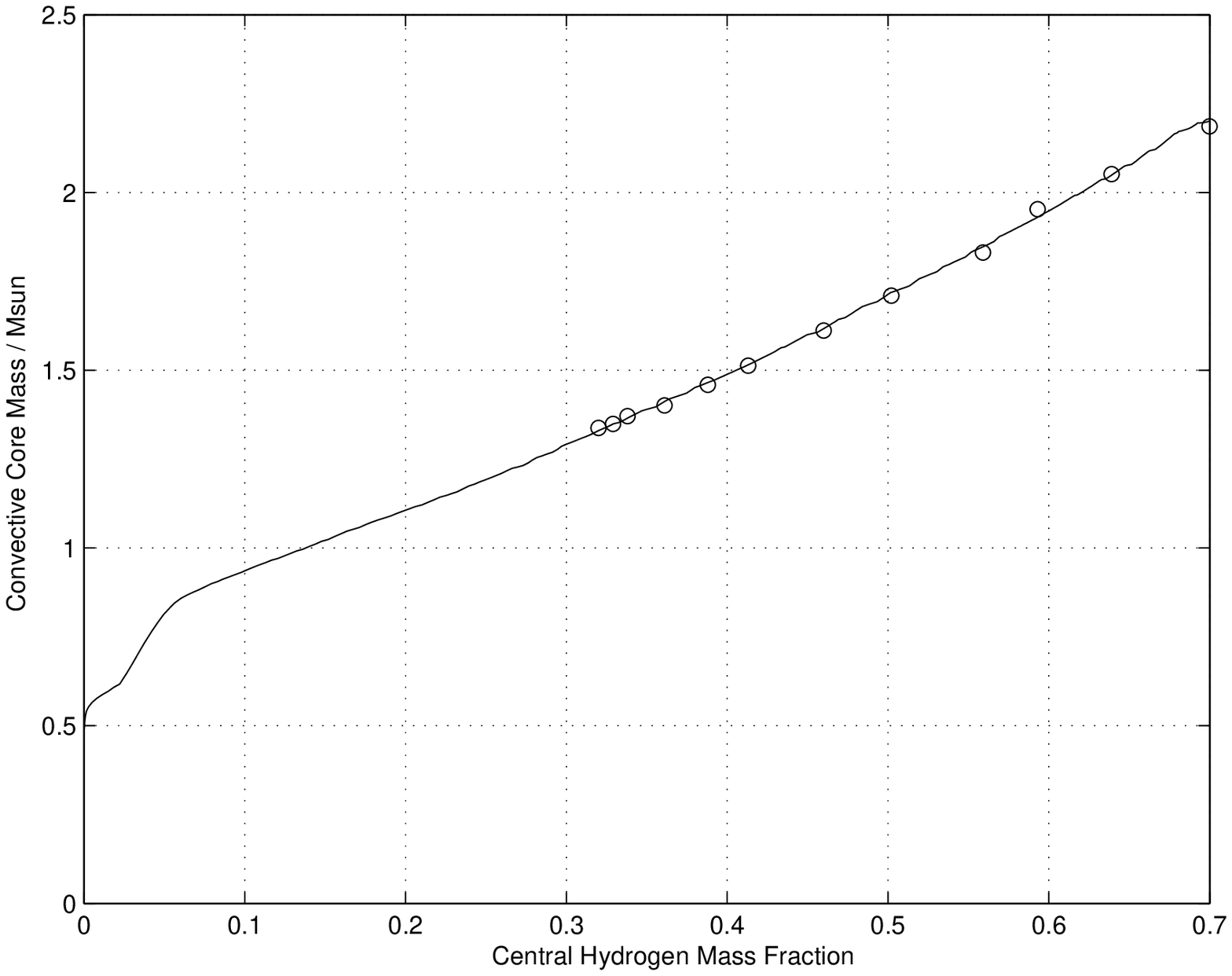}
\end{figure}

\clearpage

\begin{figure}
\plotone{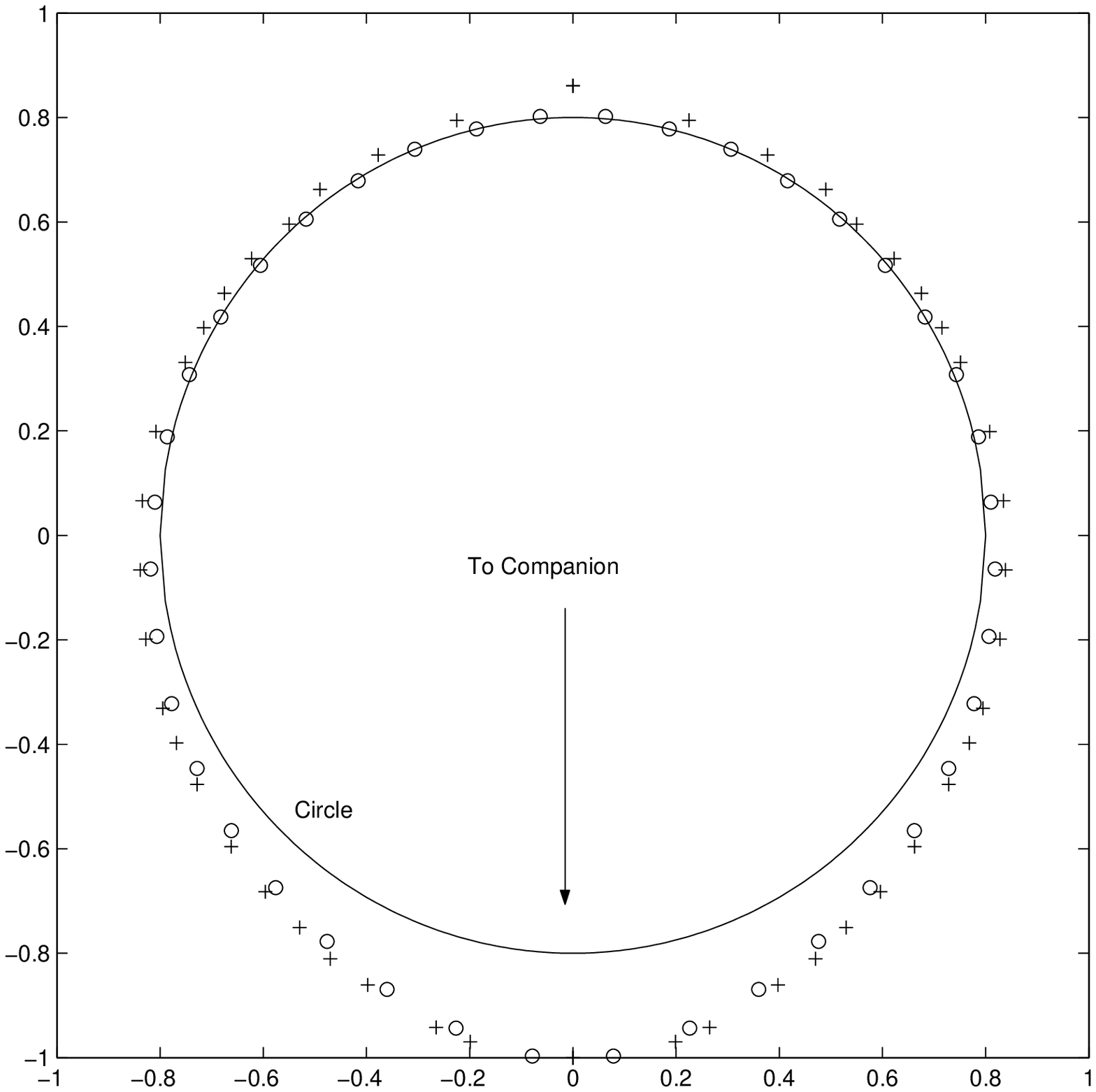}
\end{figure}

\clearpage

\begin{figure}
\plotone{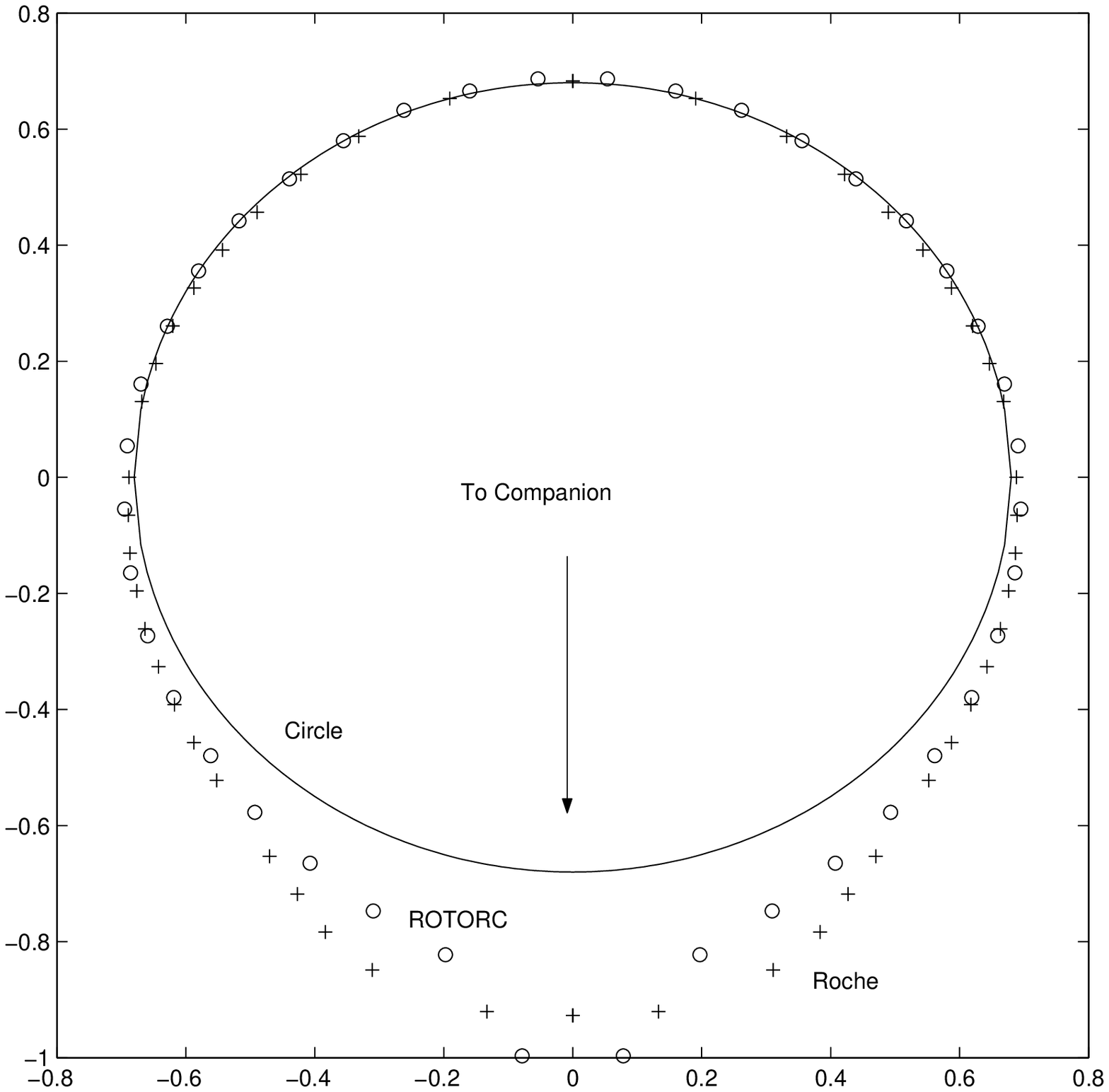}
\end{figure}

\clearpage

\begin{figure}
\plotone{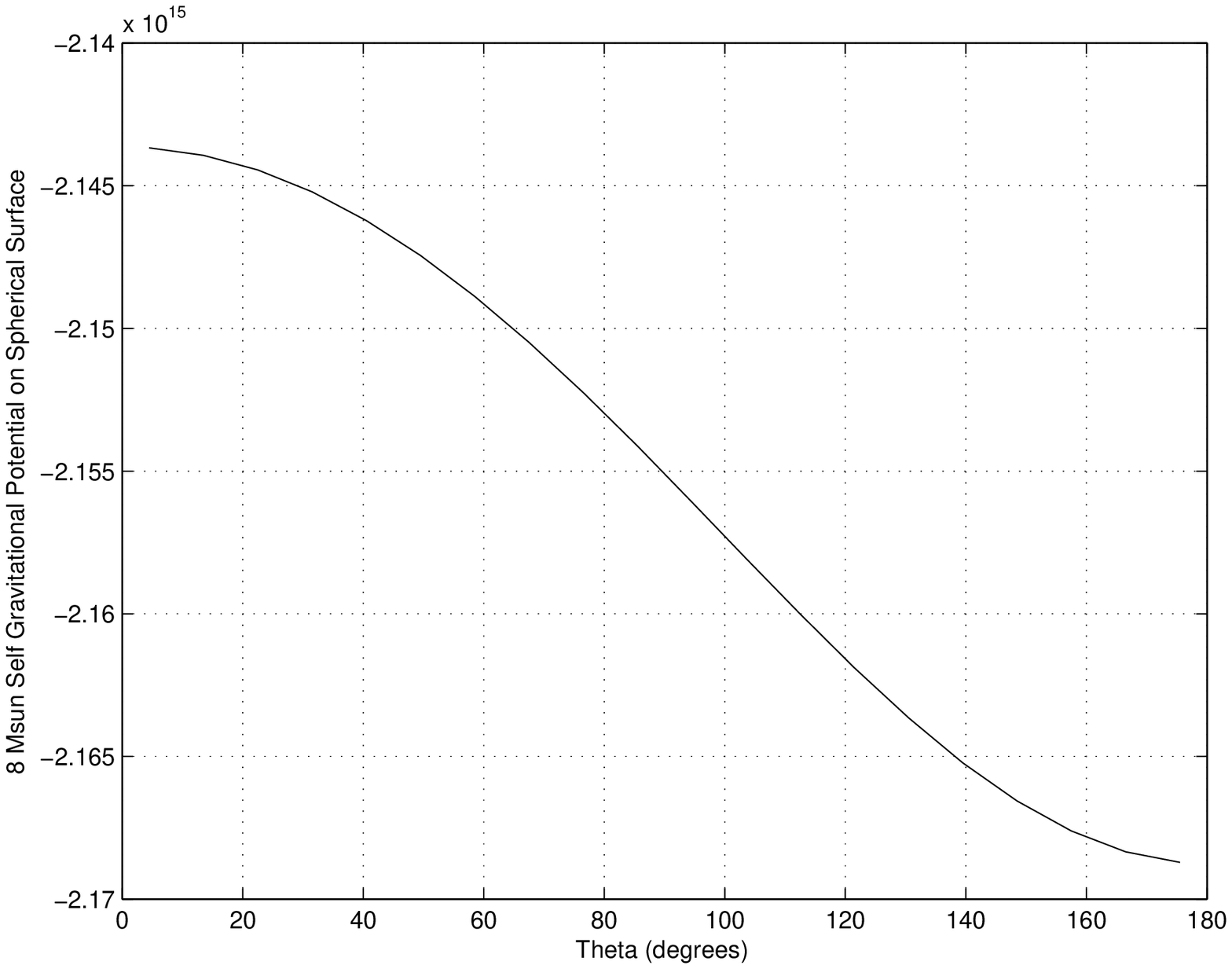}
\end{figure}

\clearpage

\begin{figure}
\plotone{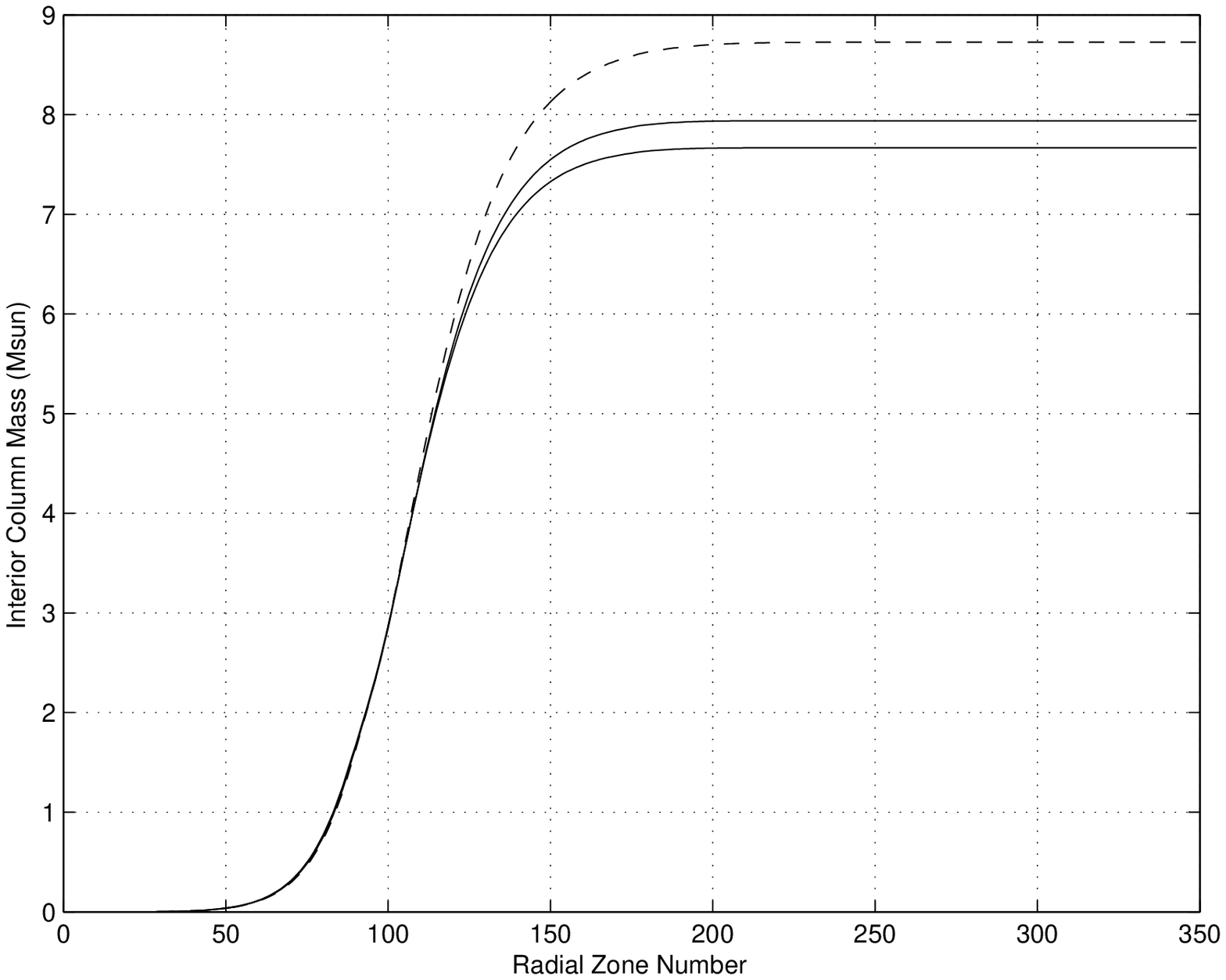}
\end{figure}

\clearpage

\begin{figure}
\plotone{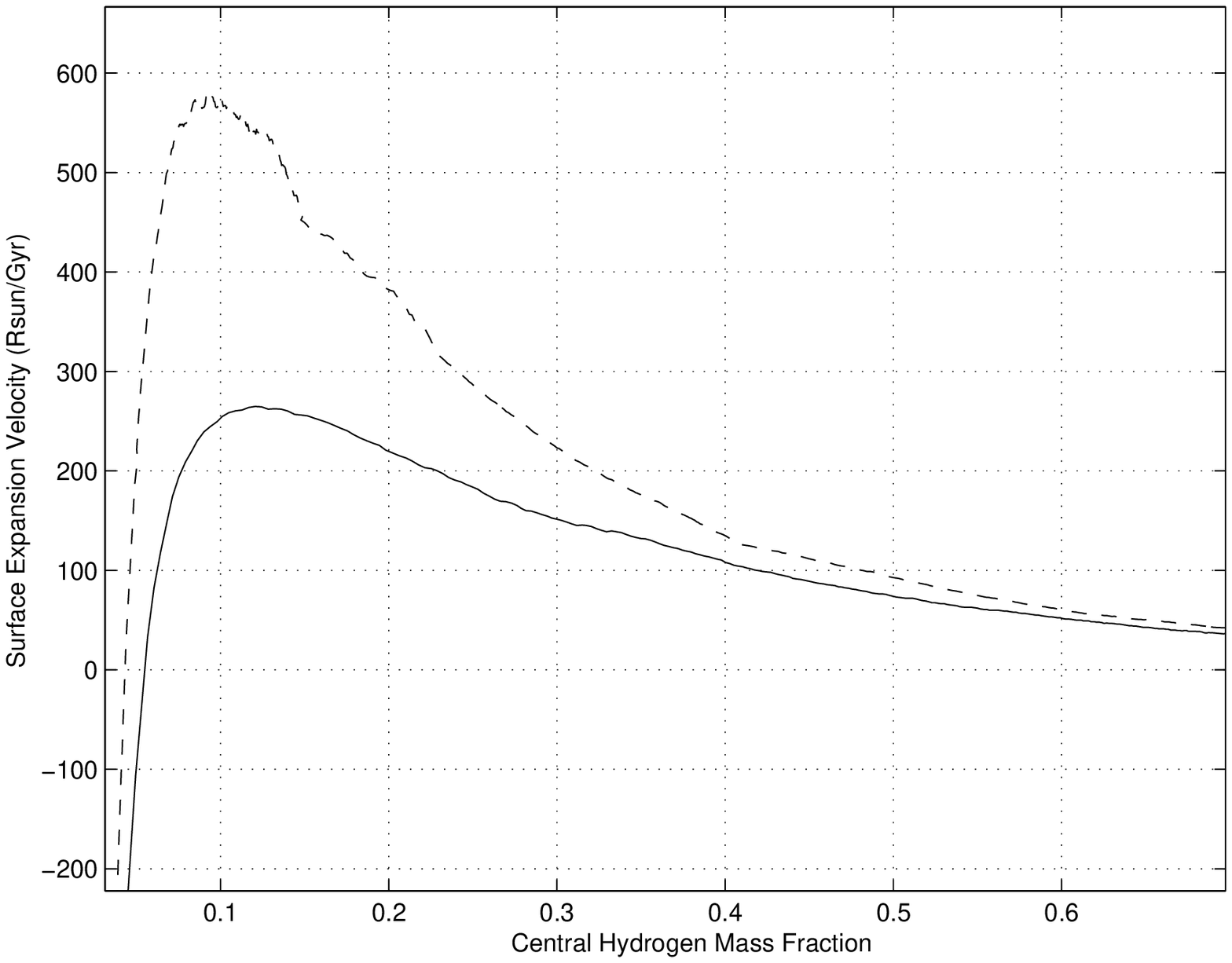}
\end{figure}

\clearpage

\begin{figure}
\plotone{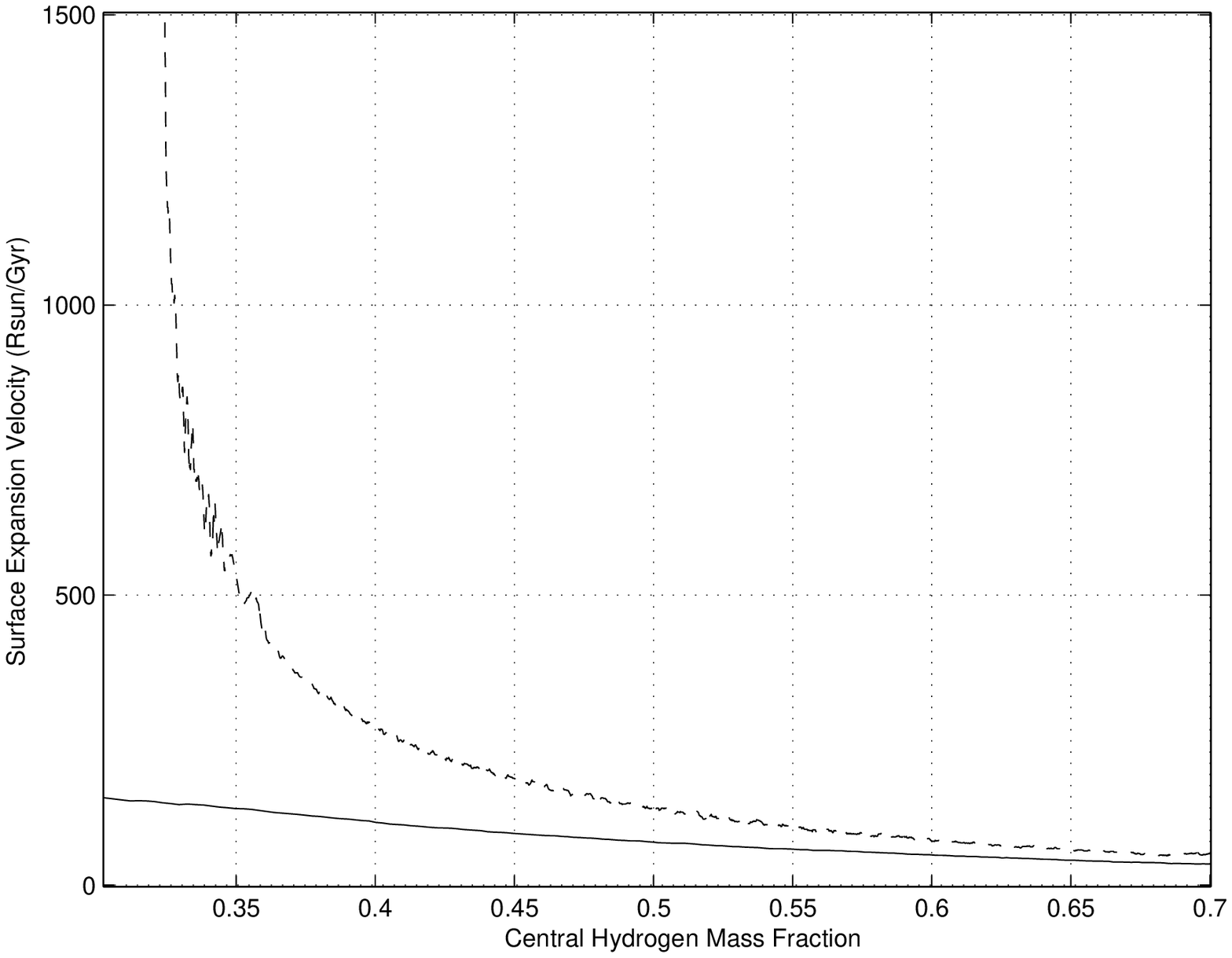}
\end{figure}

\clearpage

\begin{figure}
\plotone{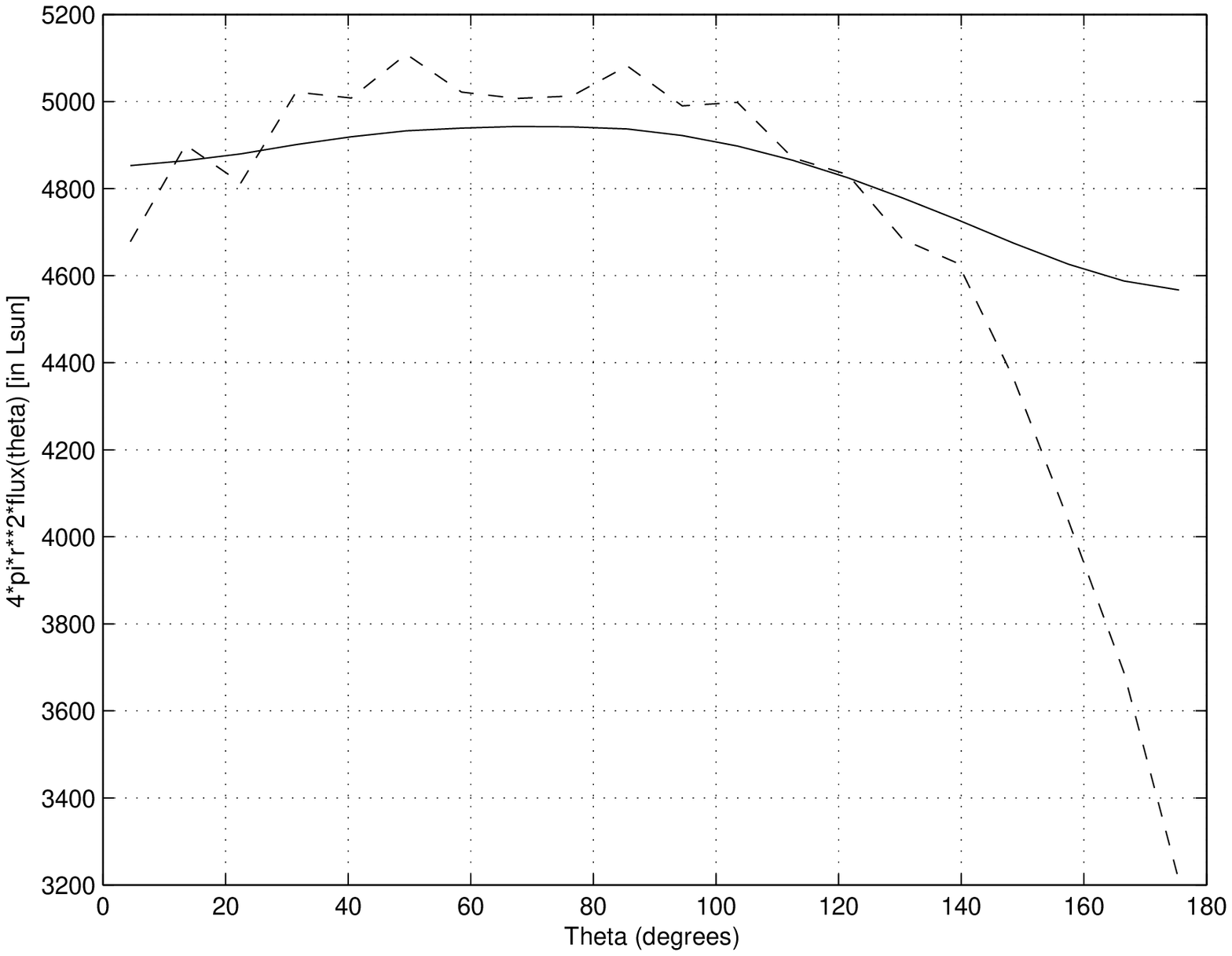}
\end{figure}

\clearpage

\begin{figure}
\plotone{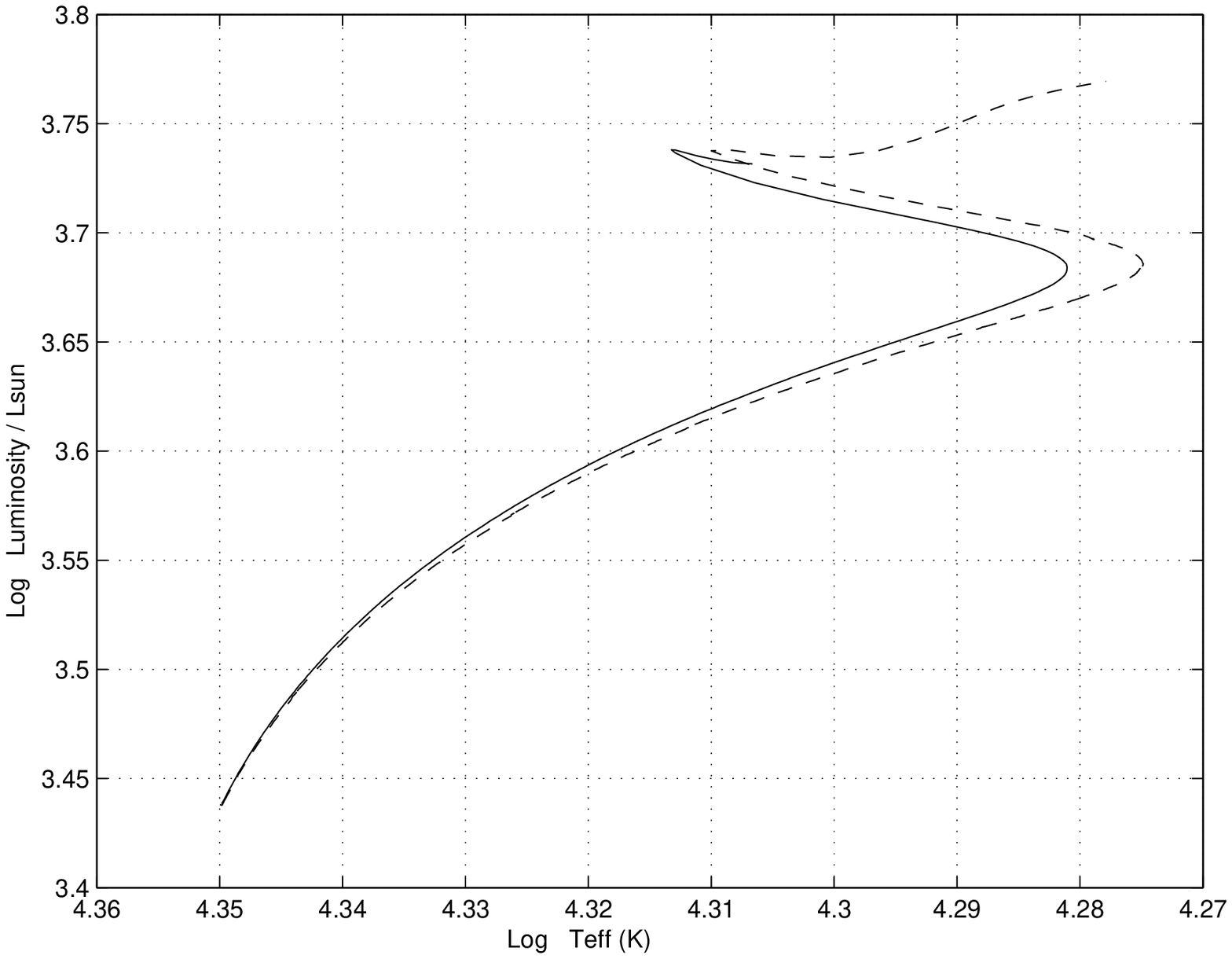}
\end{figure}





\end{document}